\documentclass[10pt,letterpaper]{article}
\ifdefined\pdfminorversion
\fi
\usepackage{spconf,amsmath,amssymb}
\usepackage[T1]{fontenc}
\usepackage{newtxtext,newtxmath}
\usepackage{graphicx,booktabs}
\usepackage[table]{xcolor}
\usepackage{cite}
\usepackage{etoolbox}
\usepackage{microtype}
\usepackage[hidelinks]{hyperref}
\usepackage{enumitem}
\usepackage{orcidlink}

\apptocmd{\thebibliography}{%
  \setlength{\itemsep}{3pt}%
  \setlength{\parsep}{0pt}%
}{}{\PackageWarning{rflocmatch-layout}{Bibliography spacing patch failed}}

\definecolor{RFRed}{HTML}{B71932}
\definecolor{RFRow}{HTML}{E8F1F7}
\definecolor{RFNote}{HTML}{F5F5F5}
\newcommand{\rfbest}[1]{%
  \textcolor{RFRed}{\textbf{#1}}%
}
\newcommand{\rfgain}[1]{%
  \textcolor{RFRed}{#1}%
}
\newcommand{\rfevaltablestyle}{%
  \renewcommand{\baselinestretch}{1}%
  \fontsize{9}{9.8}\selectfont
  \setlength{\tabcolsep}{1.2pt}%
  \renewcommand{\arraystretch}{0.85}%
  \setlength{\aboverulesep}{0.2ex}%
  \setlength{\belowrulesep}{0.25ex}%
}
\newcommand{\rftabnotestyle}{%
  \renewcommand{\baselinestretch}{1}%
  \fontsize{9}{10}\selectfont
}
\makeatletter
\AtBeginDocument{%
  \patchcmd{\@makecaption}{\vskip 10pt}{\vskip 2pt}{}%
    {\PackageWarning{rflocmatch-layout}{Caption spacing patch failed}}%
  \setlength{\floatsep}{2pt}%
  \setlength{\dblfloatsep}{2pt}%
  \setlength{\textfloatsep}{2pt plus 1pt minus 1pt}%
  \setlength{\dbltextfloatsep}{4pt}%
  \setlength{\intextsep}{2pt plus 1pt minus 1pt}%
  \setlength{\@fptop}{0pt}%
  \setlength{\@fpsep}{4pt}%
  \setlength{\@fpbot}{0pt plus 1fil}%
  \setlength{\@dblfptop}{0pt}%
  \setlength{\@dblfpsep}{4pt}%
  \setlength{\@dblfpbot}{0pt plus 1fil}%
}
\makeatother
\title{RFLocMatch: Reference-Conditioned Residual Learning from Angle--Delay Spectra for Accurate Localization}
\name{
    \small
    Leihao Zhao\,$^{1}$\orcidlink{0009-0002-9278-1758},
    Xiucheng Wang\,$^{1}$\orcidlink{0000-0003-1439-4875},
    Junxi Huang\,$^{1}$\orcidlink{0009-0000-9375-4992},
    Conghao Zhou\,$^{1}$\orcidlink{0000-0002-5727-2432},
    Yuxiang Zhang\,$^{2}$\orcidlink{0000-0003-0597-0594},
    Huixin Xu\,$^{3}$\orcidlink{0000-0001-5653-5168},
    Nan Cheng\,$^{1}$\orcidlink{0000-0001-7907-2071}
}

\address{
    \small
    \shortstack{
        $^{1}$School of Telecommunications Engineering,
        Xidian University, Xi'an 710071, China\\[2pt]
        $^{2}$State Key Laboratory of Networking and Switching Technology,
        BUPT,
        Beijing 100876, China\\[2pt]
        $^{3}$School of Electronic and Computer Engineering,
        Peking University Shenzhen Graduate School,
Shenzhen 518055, China\\[2pt]
        Corresponding author: Nan Cheng (e-mail: dr.nan.cheng@ieee.org).
    }
}
\begin{document}
\ninept
\maketitle
\flushbottom
\setlength{\abovedisplayskip}{4pt plus 1pt minus 1pt}
\setlength{\belowdisplayskip}{4pt plus 1pt minus 1pt}
\setlength{\abovedisplayshortskip}{2pt plus 1pt}
\setlength{\belowdisplayshortskip}{3pt plus 1pt minus 1pt}

\begin{abstract}
Radio fingerprint localization supports mobile robots and location-aware services when satellite positioning is unreliable. Angle--delay power fingerprints capture power distributions over angles and delays, but localization remains vulnerable to spatial ambiguity and measurement perturbations. Reflections and blockage can produce similar fingerprints at distant locations, while noise in query fingerprints can alter query--reference rankings despite a clean reference library. These effects can place a distant reference first in the ranking, creating tail errors. Similarity scores can generate plausible candidates, but reliable selection requires candidate-specific residual evidence and comparison across alternatives. This separation motivates RFLocMatch, which uses complementary joint and link-specific matching to preserve candidate locations supported by different links. Reference-conditioned residual learning then scores each candidate using the query and reference fingerprints, their absolute difference, and their element-wise product, while candidate-context modeling compares alternatives. Tail-focused training emphasizes costly errors, and inference replaces the initial match only when a candidate's score advantage exceeds a validation-calibrated threshold. On ray-tracing-generated data, RFLocMatch reduces dual-transmitter 99th-percentile error by 75.2\% under link and receiver perturbations relative to the best-performing baseline. On real-world single-transmitter channel impulse response (CIR) measurements with motion-assisted angular-spectrum estimation, it attains 8.25-m 90th-percentile error and 3.83-m mean absolute error.
\end{abstract}


\section{Introduction}
\label{sec:introduction}

Reliable localization supports mobile robots and location-aware services when satellite positioning is unreliable~\cite{liu2007survey}. Its operational value depends on routine accuracy and protection against rare errors that place a receiver in a distant region. Radio fingerprinting associates location-dependent propagation signatures with known positions~\cite{bahl2000radar,wang2017csi}, and angle--delay power fingerprints organize multipath energy across angular and delay bins without explicit path association, providing structured evidence for reference matching~\cite{hu2024adps}.

\begin{figure}[t]
    \centering
    \includegraphics[width=1.0\columnwidth]{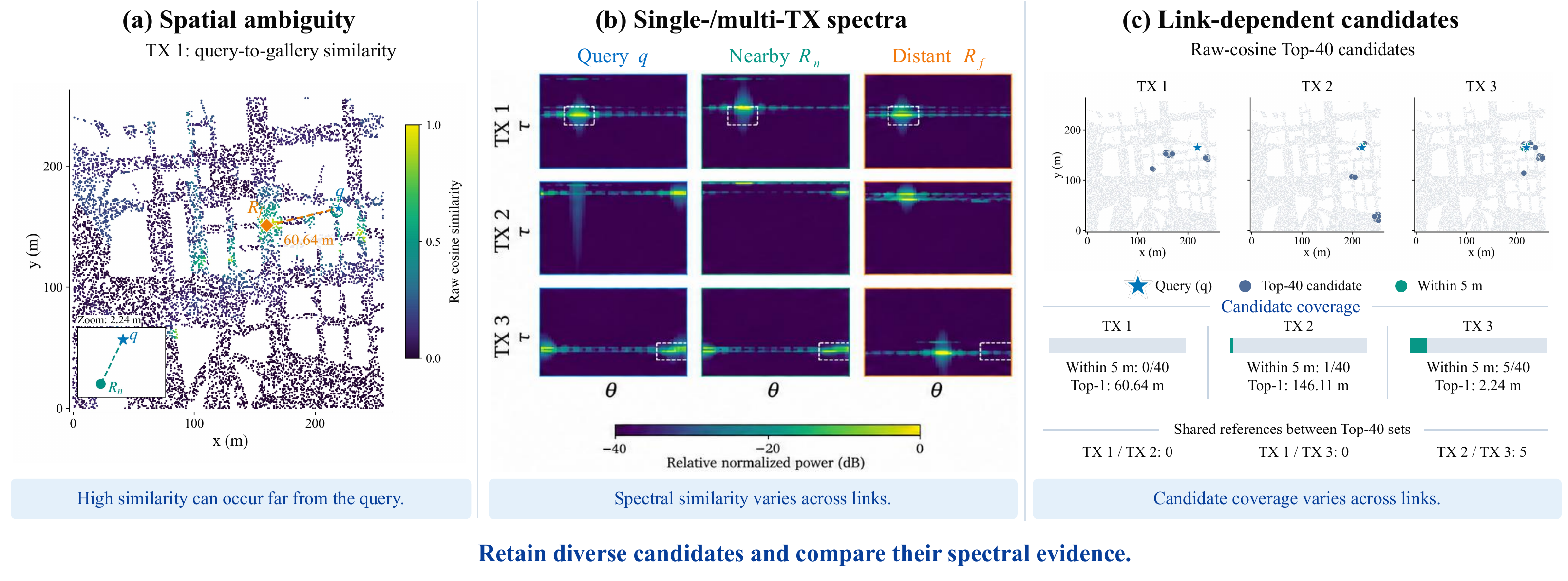}
    \caption{Simulated spatial ambiguity, link-dependent fingerprints,
    and candidate coverage.
    Proximity labels are for offline diagnosis only.}
    \label{fig:motivation}
\end{figure}

Spatial ambiguity and measurement perturbations create distinct localization challenges. Reflections and blockage allow distant locations to exhibit similar angle--delay power distributions, while individual transmitter--receiver links may favor different references. Query-side noise can alter matching scores and rankings even when the reference library remains clean. A top-ranked reference may therefore lie far from the true query location and produce a severe error (Fig.~\ref{fig:motivation}); Fig.~\ref{fig:motivation} shows the resulting ambiguity and link-complementary candidate coverage, and its distance labels diagnose retrieval ambiguity offline without entering the matcher or scorer. Since neither the true position nor the perturbation level is available online, localization requires plausible alternatives and a fingerprint-based test of whether one should replace the initial match.

Similarity matching efficiently ranks a reference library, coordinate regression captures nonlinear fingerprint mappings, and reference-assisted methods introduce nearby samples~\cite{hu2024adps,wang2017csi,cho2025madas,shah2025vit,huang2026retrieval}. However, a single joint ranking can overlook link-supported locations, while one similarity score cannot expose localized query--reference disagreement or compare evidence across candidates. Similarity can therefore generate plausible candidates, after which the absolute query--reference difference reveals mismatched energy, the element-wise product identifies aligned support, and candidate-context modeling shows whether an alternative is better supported than its competitors. This structure separates candidate generation from anchor replacement and makes the latter depend on residual evidence rather than similarity alone.

To address these issues, we propose RFLocMatch, which realizes this separation through coordinated matching and selection. Complementary reference matching interleaves joint and link-specific rankings from two learned matchers, preserving locations supported by either the complete fingerprint or an individual link. Reference-conditioned residual learning then scores each candidate using the original fingerprints, their absolute difference, their element-wise product, and matching cues, with candidate-context modeling comparing this evidence across alternatives. Tail-focused training emphasizes costly errors, while inference replaces the initial anchor only when a candidate's score advantage exceeds a validation-calibrated threshold. The framework supports separately adapted single- and multi-transmitter models without using query coordinates online. On ray-tracing data, RFLocMatch reduces dual-transmitter 99th-percentile error by 80.6\% (clean) and 75.2\% (perturbed) relative to the best-performing baseline; on real-world measured single-transmitter CIR data it attains 8.25-m 90th-percentile and 3.83-m mean absolute error (MAE). The main contributions are summarized as follows.
\begin{enumerate}[nosep]
    \item We develop complementary reference matching that unifies joint and link-specific rankings from two learned matchers, preserving locations supported globally or by individual links across separately adapted single- and multi-transmitter settings.
    \item We formulate localization as reference-conditioned candidate scoring based on the original fingerprints, absolute differences, element-wise products, retrieval cues, and candidate-context modeling.
    \item We introduce tail-focused training and a validation-calibrated anchor replacement rule to emphasize costly errors while preserving reliable initial matches.
    \item We establish the effectiveness of RFLocMatch across transmitter configurations, query perturbations, non-line-of-sight (NLOS) conditions, reference-library densities, and real-world measured single-transmitter CIRs with motion-assisted angular-spectrum estimation.
\end{enumerate}

\section{Problem Formulation}
\label{sec:problem_formulation}

Consider a receiver (RX) in $\Omega\subseteq\mathbb{R}^{2}$ observed through $U\geq1$ transmitters (TXs)~\cite{bahl2000radar,hu2024adps}, where $U=1$ and $U\geq2$ denote single- and multi-TX localization. Transmitter $u$ provides an angle--delay power fingerprint $\mathbf{P}_u\in\mathbb{R}_{+}^{B_{\tau}\times B_{\theta}}$, the receiver fingerprint is $\mathcal{F}=(\mathbf{P}_1,\ldots,\mathbf{P}_U)$, and the reference library stores $N$ fingerprints with known positions. Query and reference fingerprints share transmitter identities and binning; only the query may be perturbed. Given $\mathcal{Q}$ at unknown position $\mathbf{x}^{*}$, a localization rule $f$ produces
\begin{equation}
    \hat{\mathbf{x}}=f(\mathcal{Q},\mathcal{D}),
    \qquad
    e=\|\hat{\mathbf{x}}-\mathbf{x}^{*}\|_2,
    \label{eq:localization_problem}
\end{equation}
where $e$ is the Euclidean localization error.

\section{Proposed Method}
\label{sec:framework}

\begin{figure}[!tb]
    \centering
    \includegraphics[width=1.0\columnwidth]{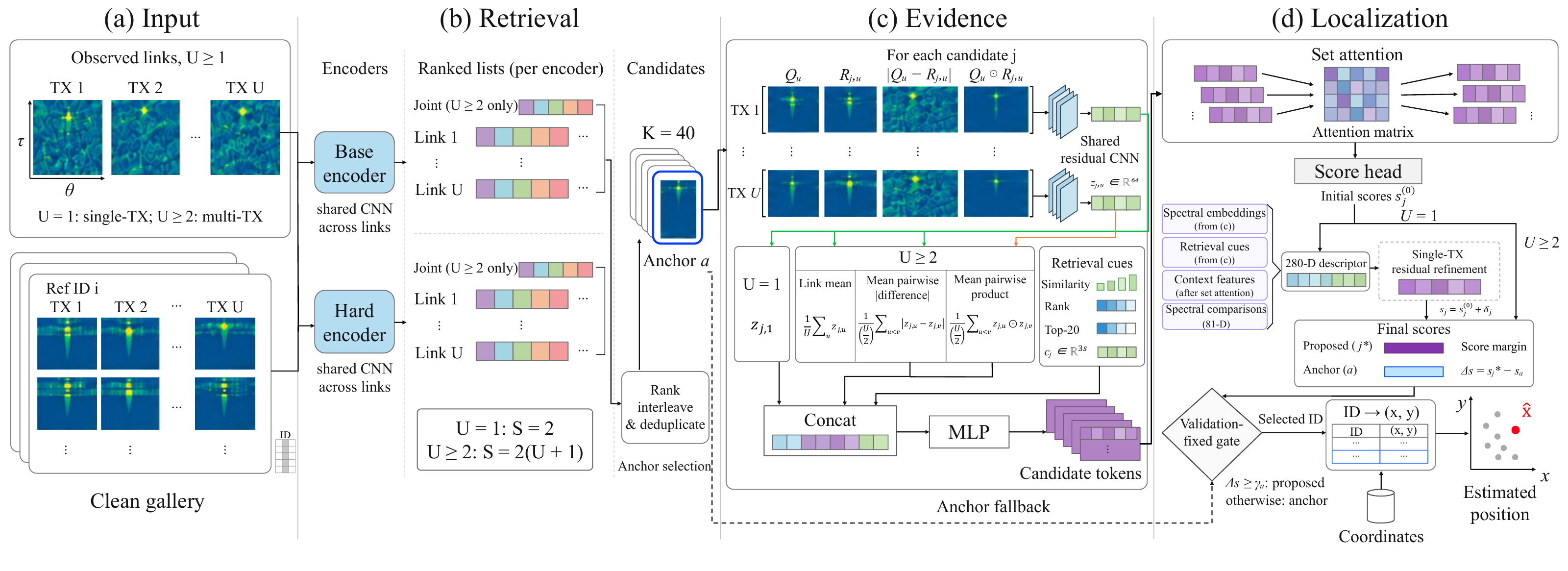}
    \caption{RFLocMatch pipeline for (a) preprocessing, (b) matching, (c) residual evidence, and (d) candidate scoring with auxiliary context and gating.}
    \label{fig:rflocmatch_framework}
\end{figure}

The RFLocMatch pipeline is shown in Fig.~\ref{fig:rflocmatch_framework} for separately trained single- and multi-TX models; fingerprints are aligned by transmitter identity, normalized by total power, square-rooted element-wise and $\ell_2$-normalized~\cite{arandjelovic2012rootsift}.

Complementary matchers retain a candidate set and an initial anchor; a validation-calibrated gate limits anchor replacement in simulation, whereas the real-world configuration keeps the same stages and aggregates the top candidates. Graph attention network (GAT) methods regress coordinates from graphs of retrieved channel state information (CSI) references~\cite{huang2026retrieval}. RFLocMatch differs in three respects: retrieved references remain discrete hypotheses; the scorer is conditioned on the query--reference pair through residual and agreement evidence; and reference coordinates are hidden until after selection.

\subsection{Complementary Reference Matching}

The base matcher captures general similarity; the hard matcher emphasizes confusers. Both use a link-shared convolutional neural network (CNN) that aggregates link embeddings as
\begin{equation}
    G_U(\{\mathbf{v}_u\}_{u=1}^{U})=
    \begin{cases}
        \mathbf{v}_1, & U=1,\\
        [\bar{\mathbf{v}},\boldsymbol{\delta},
        \boldsymbol{\pi}], & U\geq2,
    \end{cases}
    \label{eq:link_fusion}
\end{equation}
Here brackets denote concatenation, $\bar{\mathbf{v}}$ is the link mean, $\boldsymbol{\delta}$ and $\boldsymbol{\pi}$ average link differences and products, and the embedding is order-invariant~\cite{zaheer2017deepsets}, normalized directly for $U=1$ and projected otherwise.
Each matcher returns one joint ranking and, for $U\geq2$, $U$ link-specific rankings; interleaving and deduplicating them yields a candidate set $\mathcal{C}$ of size $K=40$, with anchor $a$ maximizing summed similarities within a separate Top-$K$ pool. The cue $\mathbf{c}_j\in\mathbb{R}^{3S}$ records each source's similarity, rank in $\mathcal{C}$ and Top-$20$ membership. Complementary rankings serve candidate preservation rather than localization.

\subsection{Reference-Conditioned Residual Learning}

The scorer performs hypothesis verification, not coordinate regression; reference coordinates are unavailable throughout scoring. Since global scores can conceal localized disagreements, an evidence tensor is built for candidate $j$ and link $u$ from the preprocessed fingerprints $\mathbf{Q}_u$ and $\mathbf{R}_{j,u}$:
\[
    \mathbf{E}_{j,u}
    =\operatorname{Stack}\!\left(
        \mathbf{Q}_u,\mathbf{R}_{j,u},
        |\mathbf{Q}_u-\mathbf{R}_{j,u}|,
        \mathbf{Q}_u\odot\mathbf{R}_{j,u}
    \right).
\]
The residual exposes disagreement that a high global similarity can hide, while the product highlights regions supported by both.
A shared residual CNN maps $\mathbf{E}_{j,u}$ to $\mathbf{z}_{j,u}\in\mathbb{R}^{64}$, and a multilayer perceptron (MLP) $\phi_U$ fuses the aggregated evidence with the retrieval cues as
\[
    \mathbf{h}_j=\phi_U\!\left(
        [G_U(\{\mathbf{z}_{j,u}\}_{u=1}^{U}),\mathbf{c}_j]
    \right).
\]
Candidates are scored jointly rather than independently. A one-layer, four-head Transformer~\cite{vaswani2017attention,lee2019settransformer} compares candidate tokens, and layer normalization~\cite{ba2016layernorm} with a linear head yields initial scores $s_j^{(0)}$. For multi-TX localization, cross-link disagreement provides a consistency cue; since this signal is unavailable for single TX, we introduce a compact auxiliary spectral-agreement descriptor only for single-TX refinement, summarizing full-spectrum, angle-marginal, delay-marginal and multiscale agreement (4, 4, 4 and 69 terms); it is not used in the multi-TX path. Reference coordinates are excluded from the scoring function and accessed only after candidate selection. With the evidence encoding, the cues and $s_j^{(0)}$, it forms the input to a residual MLP--Transformer head whose zero-initialized output layer yields $\delta_j$, so $s_j=s_j^{(0)}+\delta_j$ for $U=1$ and $s_j=s_j^{(0)}$ for $U\geq2$. As an auxiliary safeguard in simulation, a validation-calibrated gate outputs $\mathbf{x}_{j^\star}$ for $j^\star=\arg\max_{j\in\mathcal{C}}s_j$ if $\Delta s=s_{j^\star}-s_a\geq\gamma_U$, and $\mathbf{x}_a$ otherwise, with $\gamma_U$ validated before testing. Scoring is coordinate-free: reference coordinates are read only after selection.

\subsection{Tail-Focused Training}

Both matchers use bidirectional contrastive cross-entropy (CE)~\cite{radford2021clip}; the hard matcher adds a hinge penalty against the most similar in-batch reference at least 40~m away.
The scorer uses distance-aware targets $d_{ij}=\|\mathbf{x}_j-\mathbf{x}_i^\star\|_2$, proportional to $\exp(-d_{ij}/2.5)$ for $d_{ij}\leq10$~m and concentrated on the nearest candidate otherwise. For $p_{ij}=\operatorname{softmax}_j(s_{ij})$, the clipped risk and objective are
\begin{equation}
\begin{aligned}
r_i&=\sum_{j\in\mathcal{C}_i}p_{ij}
\frac{\min(d_{ij},d_{\max})}{d_{\max}},\\
\mathcal{L}_U&=\mathcal{L}_{\mathrm{CE}}+\mathcal{L}_{\mathrm{margin}}
+\lambda_U\operatorname{TopMean}_{25\%}(\{r_i\})\\
&\quad+\mathcal{L}_{\mathrm{reg},U}.
\end{aligned}
    \label{eq:rf_loss}
\end{equation}
Here $d_{\max}=150$~m, $\mathcal{L}_{\mathrm{CE}}$ fits the normalized distance targets, and $\mathcal{L}_{\mathrm{margin}}$ penalizes top-scoring candidates at $\leq5$~m and $\geq40$~m. $\operatorname{TopMean}_{25\%}$ averages the largest quarter of batch risks~\cite{fan2017topk}, with $\lambda_U=0.5$ in base training and $2$ in single-TX refinement.
The tail-aware term acts as a training regularizer, secondary to reference-conditioned rescoring. $\mathcal{L}_{\mathrm{reg},1}$ adds twice the batch-mean mass beyond 40~m with Kullback--Leibler (KL) distillation from ungated teacher errors $\leq5$~m~\cite{hinton2015distilling}; the residual head trains with the base scorer frozen.

\section{Experimental Evaluation}
\label{sec:experiments}

\subsection{Experimental Setup}
\label{sec:experimental_setup}

The D$^2$LoS map~\cite{wang2026d2los} has 13,877 RX locations, ten TXs and a $51\times45$ angle--delay fingerprint per link. The receiver split uses 10,000 locations for training and gallery, 500 for selection, 1,377 for confirmation and 2,000 for testing; five folds partition the 45 dual-TX pairs into 30/6/9 splits. Baselines are ADPS-AWKNN~\cite{hu2024adps}, radio-only MADAS~\cite{cho2025madas}, ViT-UE~\cite{shah2025vit} and ADPS-GAT, all trained on the same split, queries, gallery and perturbations with hyperparameters selected on validation. ADPS-AWKNN is the strongest native angle--delay fingerprint baseline and serves as the primary reference for comparison; MADAS, ViT-UE and ADPS-GAT are adapted under the same representation and protocol as controlled architectural comparisons, replacing only the unavailable input encoder, so ADPS-GAT keeps its published retrieval, graph, GAT and localization heads with the ADPS feature substituted for the CSI-chart node encoder. The 1,377 confirmation locations are a post-freeze sanity check only.
Dual-TX query perturbations are applied before normalization as
\[
\widetilde{\mathbf{P}}_{q,u}
=
\left[
\mathbf{P}_{q,u}
+\rho_{q,u}10^{-L/20}\mathbf{Z}^{\ell}_{q,u}
+\sqrt{A}\,10^{-R/20}\mathbf{Z}^{r}_{q,u}
\right]_+,
\]
with $L$, $R$ the link and receiver perturbation strengths, $\rho_{q,u}^{2}$ scaling link noise by clean query power, and $A$ the gallery median. The 30 ordered $L\ne R$ pairs from $\{\infty,15,12.5,10,7.5,5\}$~dB use three fixed noise realizations and indicate perturbation strength, not a calibrated signal-to-noise ratio (SNR). Ablations compare anchor-only (Anc.), joint-link-only (Pair), cues-only (Cues) and independent (Ind.) scoring. The dual-NLOS evaluation covers 50,682 clean queries with both TX links NLOS, and galleries of 2,500--7,500 locations are tested without retraining. We report the 90th, 95th and 99th percentiles (P90, P95, P99) and MAE, macro-averaged per TX group, with P99 intervals from 5,000 paired spatial-block bootstrap resamples.

For the real-world campaign, complex CIRs and LiDAR relative motion from WiWorld-RealData~\cite{jiao2026wiworld} estimate angular power. Within each half-second window we select up to six delay peaks and fit candidate directions per peak on a $1^\circ$ grid with local refinement:
\begin{equation}
\begin{aligned}
(\hat{\boldsymbol\theta}_d,\hat{\mathbf B}_d)&=\arg\min_{\boldsymbol\theta,\mathbf B}\|\mathbf H_d-\mathbf S(\boldsymbol\theta)\mathbf B\|_F^2,\\
S_{n\ell}(\boldsymbol\theta)&=\exp\!\left(\mathrm{i}\frac{2\pi f_c}{c}\Delta\mathbf p_n^\top\mathbf u(\theta_\ell)\right),
\end{aligned}
\label{eq:estimated_aps}
\end{equation}
Here $\mathbf H_d$ is the complex CIR around delay peak $d$, $\Delta\mathbf p_n$ is LiDAR-estimated displacement, $\mathbf u(\theta)=[\cos\theta,\sin\theta]^\top$, and $f_c=3.7$~GHz. Candidate directions are selected by reconstruction on the held-out 0.3--0.4~s portion and refit on 0--0.4~s, disjoint from the final 0.4--0.5~s; the spectra are rotated into the LiDAR body frame and smoothed into a normalized 48-bin angular power spectrum (APS), with a 128-bin power delay profile (PDP) from the same CIR. LiDAR supplies only relative motion for angular-spectrum estimation, and no LiDAR position, map coordinate or localization label enters scoring or inference. A one-dimensional CNN regressor on the same inputs reaches 41.2-m P90 and 18.6-m MAE, so the gain is not a representation artifact; baselines needing a two-dimensional representation receive a PDP$\times$APS proxy from the same real-world measured PDP. Of 551 label-eligible half-second windows, 330/110/111 form training/gallery, validation and test at group level, and the real-world campaign is single-transmitter, so $K=20$.

\subsection{Main Performance Comparison}
\label{sec:main_performance}

\begin{figure}[!tb]
  \centering
  \includegraphics[width=1.0\columnwidth]{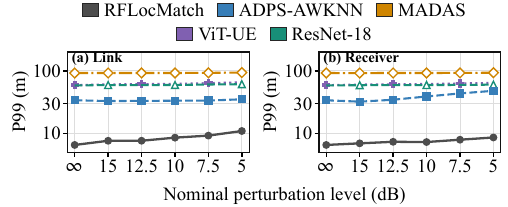}
  \caption{P99 under isolated (a) link perturbation and (b) receiver noise on a logarithmic scale ($\infty$: Clean).}
  \label{fig:p99_component}
\end{figure}

RFLocMatch reduces dual-TX P99 by 80.6\% (clean) and 75.2\% (perturbed) relative to ADPS-AWKNN (Table~\ref{tab:dual_tx_main}), with 95\% intervals of 71.6--83.8\% and 66.2--77.6\%. The improvement is concentrated in the severe-error tail: P90 changes are modest while P99 decreases sharply. P99 stays below 11~m across the perturbation sweeps (Fig.~\ref{fig:p99_component}), and single- and three-TX P99 falls by 47.5\% and 52.7\% (Table~\ref{tab:tx_count_clean}), with the dual-TX setting showing the largest relative reduction and the adapted baselines staying above 59~m P99 in both query conditions. 

\begin{table}[!t]
\centering
\caption{Dual-TX localization under clean and noisy queries.}
\label{tab:dual_tx_main}

\rfevaltablestyle

\begin{tabular}{llrrrr}
\toprule
Query & Method
& P90$\downarrow$ & P95$\downarrow$
& P99$\downarrow$ & MAE$\downarrow$ \\
\midrule

Clean & ADPS-AWKNN~\cite{hu2024adps}
& \underline{3.23} & \underline{4.33}
& \underline{33.91} & \underline{2.81} \\

& MADAS$^\dagger$~\cite{cho2025madas}
& 37.44 & 52.54 & 92.24 & 17.87 \\

& ViT-UE$^\dagger$~\cite{shah2025vit}
& 22.02 & 31.58 & 59.00 & 10.10 \\

& ADPS-GAT$^\dagger$~\cite{hu2024adps,huang2026retrieval}
& 40.54 & 61.21 & 106.70 & 15.16 \\

\rowcolor{RFRow}
& \textbf{RFLocMatch}
& \rfbest{3.14} & \rfbest{3.83}
& \rfbest{6.57} & \rfbest{2.14} \\

\midrule
\rowcolor{RFNote}
\multicolumn{2}{l}{\textbf{Reduction}}
& \rfgain{2.9\%} & \rfgain{11.6\%}
& \rfgain{80.6\%} & \rfgain{23.6\%} \\

\midrule

Noisy & ADPS-AWKNN~\cite{hu2024adps}
& \underline{3.41} & \underline{4.55}
& \underline{37.29} & \underline{2.92} \\

& MADAS$^\dagger$~\cite{cho2025madas}
& 37.78 & 53.15 & 93.31 & 18.03 \\

& ViT-UE$^\dagger$~\cite{shah2025vit}
& 24.44 & 34.79 & 63.73 & 11.28 \\

& ADPS-GAT$^\dagger$~\cite{hu2024adps,huang2026retrieval}
& 39.96 & 59.58 & 105.55 & 14.81 \\

\rowcolor{RFRow}
& \textbf{RFLocMatch}
& \rfbest{3.25} & \rfbest{4.09}
& \rfbest{9.24} & \rfbest{2.31} \\

\midrule
\rowcolor{RFNote}
\multicolumn{2}{l}{\textbf{Reduction}}
& \rfgain{4.7\%} & \rfgain{10.1\%}
& \rfgain{75.2\%} & \rfgain{20.9\%} \\

\bottomrule
\end{tabular}

\par\vspace{1pt}
\parbox{\linewidth}{\rftabnotestyle\raggedright
Errors in meters; the dagger marks power-fingerprint implementations, and ADPS-GAT shares the ADPS-AWKNN candidate set.\par}

\end{table}

\begin{table}[!t]
\centering
\caption{Clean single-TX and three-TX localization.}
\label{tab:tx_count_clean}

\rfevaltablestyle

\begin{tabular}{llrrrr}
\toprule
TXs & Method
& P90$\downarrow$ & P95$\downarrow$
& P99$\downarrow$ & MAE$\downarrow$ \\
\midrule

1 & ADPS-AWKNN~\cite{hu2024adps}
& \underline{4.77} & \underline{25.99}
& 129.25 & \underline{6.45} \\

& MADAS$^\dagger$~\cite{cho2025madas}
& 59.09 & 81.04 & 133.94 & 26.23 \\

& ViT-UE$^\dagger$~\cite{shah2025vit}
& 46.52 & 63.88 & \underline{104.29} & 19.68 \\

& ADPS-GAT$^\dagger$~\cite{hu2024adps,huang2026retrieval}
& 60.85 & 83.67 & 124.01 & 23.11 \\

\rowcolor{RFRow}
& \textbf{RFLocMatch}
& \rfbest{3.87} & \rfbest{5.36}
& \rfbest{54.78} & \rfbest{3.55} \\

\midrule
\rowcolor{RFNote}
\multicolumn{2}{l}{\textbf{Reduction}}
& \rfgain{18.8\%} & \rfgain{79.4\%}
& \rfgain{47.5\%} & \rfgain{45.0\%} \\

\midrule

3 & ADPS-AWKNN~\cite{hu2024adps}
& \underline{3.11} & \underline{3.97}
& \underline{10.53} & \underline{2.28} \\

& MADAS$^\dagger$~\cite{cho2025madas}
& 26.40 & 35.79 & 66.69 & 13.12 \\

& ViT-UE$^\dagger$~\cite{shah2025vit}
& 12.45 & 17.78 & 37.25 & 6.39 \\

& ADPS-GAT$^\dagger$~\cite{hu2024adps,huang2026retrieval}
& 31.59 & 49.98 & 92.86 & 11.48 \\

\rowcolor{RFRow}
& \textbf{RFLocMatch}
& \rfbest{3.00} & \rfbest{3.21}
& \rfbest{4.98} & \rfbest{1.85} \\

\midrule
\rowcolor{RFNote}
\multicolumn{2}{l}{\textbf{Reduction}}
& \rfgain{3.6\%} & \rfgain{19.1\%}
& \rfgain{52.7\%} & \rfgain{19.0\%} \\

\bottomrule
\end{tabular}
\end{table}

\subsection{Dual-NLOS Analysis}

Table~\ref{tab:dual_nlos} isolates clean dual-NLOS queries, where RFLocMatch reduces P99 from 64.82 to 16.08~m and MAE from 3.61 to 2.38~m.
\begin{table}[!t]
\centering
\caption{Clean localization on the dual-NLOS subset.}
\label{tab:dual_nlos}

\rfevaltablestyle

\begin{tabular}{lrrrr}
\toprule
Method
& P90$\downarrow$ & P95$\downarrow$
& P99$\downarrow$ & MAE$\downarrow$ \\
\midrule

ADPS-AWKNN~\cite{hu2024adps}
& \underline{3.62} & \underline{6.57}
& \underline{64.82} & \underline{3.61} \\

MADAS$^\dagger$~\cite{cho2025madas}
& 47.94 & 64.77 & 104.54 & 22.69 \\

ViT-UE$^\dagger$~\cite{shah2025vit}
& 28.41 & 39.07 & 67.65 & 13.44 \\

ADPS-GAT$^\dagger$~\cite{hu2024adps,huang2026retrieval}
& 55.03 & 75.08 & 119.18 & 22.06 \\

\rowcolor{RFRow}
\textbf{RFLocMatch}
& \rfbest{3.18} & \rfbest{4.03}
& \rfbest{16.08} & \rfbest{2.38} \\

\midrule
\rowcolor{RFNote}
\textbf{Reduction}
& \rfgain{12.2\%} & \rfgain{38.7\%}
& \rfgain{75.2\%} & \rfgain{34.2\%} \\

\bottomrule
\end{tabular}

\end{table}

\subsection{Effect of Reference-Library Size}

\begin{figure}[!tb]
    \centering
    \includegraphics[width=1.0\columnwidth]
    {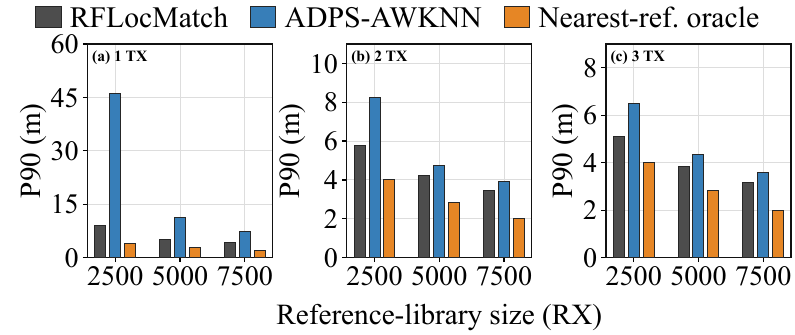}
    \caption{Clean P90 versus library size for (a) single-, (b) dual-, and (c) three-TX settings.}
    \label{fig:library_p90}

    \vspace{0pt}
    \includegraphics[width=1.0\columnwidth]
    {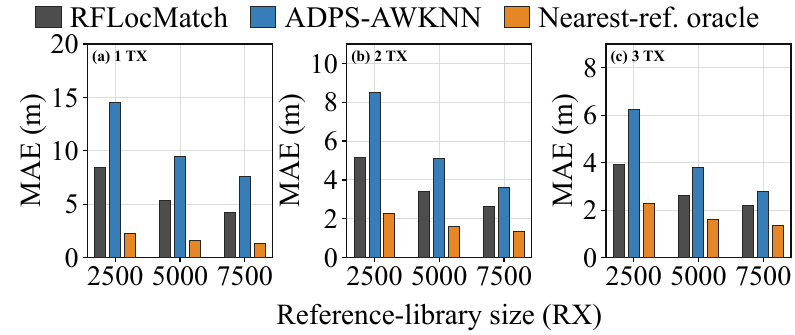}
    \caption{Clean MAE versus library size for (a) single-, (b) dual-, and (c) three-TX settings.}
    \label{fig:library_mae}

    \vspace{0pt}
    \includegraphics[width=72mm]
    {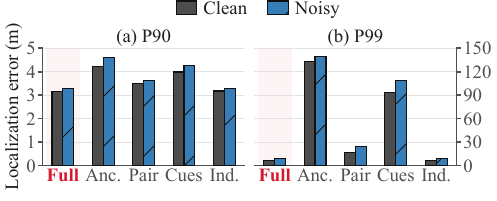}
    \caption{Dual-TX ablation for (a) P90 and (b) P99.}
    \label{fig:structure_ablation}
\end{figure}

Figs.~\ref{fig:library_p90} and~\ref{fig:library_mae} show that RFLocMatch remains better than ADPS-AWKNN across all library sizes and TX configurations; at 2,500 single-TX references P90 and MAE fall by 80.4\% and 41.6\%. The P90 reduction exceeds the MAE reduction at this operating point, matching the tail-focused behavior of the main results, and the separation persists as the gallery grows to 7,500 references, where more distractors enter the candidate pool; absolute errors are smaller in the multi-TX panels, where several links supply complementary candidates.

\subsection{Ablation Study}

A validation-selected weighted k-nearest-neighbor readout on the same candidate sets still leaves P99 at 126.7~m, whereas reference-conditioned rescoring reaches single-digit metres. Fig.~\ref{fig:structure_ablation} reports matched runs with a different control set; an offline oracle yields P99 of 3.21/3.29~m, Anc. 133.06/139.28~m, ungated Full 6.22/9.47~m and Full 6.37/9.31~m, so rescoring drives the gain and gating is auxiliary. Ind.\ scores each candidate independently without candidate-set interaction, showing that verification benefits from comparison among competing hypotheses. Removing the absolute difference or the product raises P99 to 6.74/8.09~m (clean) and 10.25/10.00~m (perturbed); removing the top-quartile risk term leaves 6.51/8.35~m. Retrieval therefore provides plausible candidates, but reliable localization requires verifying them.

\subsection{Real-World Measurements with Motion-Assisted Angular Spectra}
\label{sec:measured_diagnostic}

Table~\ref{tab:measured_estimated_aps} compares single-TX localization on real-world measured CIRs, applied with $K=20$; receiver positions come from global navigation satellite system (GNSS) time labels and are used only for offline scoring. LiDAR contributes only relative motion and no label reaches the matcher or scorer, so this is not a radio-only angular-estimation claim. The readout averages the top three scored candidates with $k=3$ fixed on validation. It attains 8.25-m P90 and 3.83-m MAE, 72.0\% and 70.9\% below ViT-UE; tail errors above 20~m occur for 1.8\% of test windows against 16.2\% for the anchor. The gain appears before coordinate aggregation: the retrieval anchor gives 69.2-m P90, whereas the highest-scored single reference already reaches 7.96~m, so it comes from hypothesis verification rather than averaging.    

\begin{table}[!t]
\centering
\caption{Single-TX real-world measured localization with estimated APS and PDP (m).}
\label{tab:measured_estimated_aps}
\rfevaltablestyle
\begin{tabular}{lrr}
\toprule
Method & P90$\downarrow$ & MAE$\downarrow$ \\
\midrule
ADPS-AWKNN~\cite{hu2024adps} & 71.64 & 22.31 \\
MADAS$^\dagger$~\cite{cho2025madas} & 39.83 & 16.87 \\
ViT-UE$^\dagger$~\cite{shah2025vit} & 29.48 & 13.15 \\
ADPS-GAT$^\dagger$~\cite{hu2024adps,huang2026retrieval} & 40.78 & 19.94 \\
\rowcolor{RFRow}
\textbf{RFLocMatch} & \rfbest{8.25} & \rfbest{3.83} \\
\midrule
\rowcolor{RFNote}
\textbf{Reduction} & \rfgain{72.0\%} & \rfgain{70.9\%} \\
\bottomrule
\end{tabular}
\par\vspace{1pt}
\parbox{\linewidth}{\rftabnotestyle\raggedright
$^\dagger$PDP$\times$APS proxy from the same real-world measured PDP; reduction is relative to ViT-UE.\par}
\end{table}

\section{Conclusion}

RFLocMatch reduces localization tail errors by rescoring retrieved references. This is the dominant gain in ablations, lowering dual-transmitter P99 by 80.6\% and 75.2\% with a small candidate set. On GNSS-labeled, real-world single-transmitter measurements with motion-assisted angular spectra, it attains 8.25-m P90 and 3.83-m MAE. Multi-transmitter field evaluation remains future work.

\section{Acknowledgment}
The authors used generative AI to translate their Chinese draft into English throughout this paper. They reviewed the translation and remain responsible for the final text.

\clearpage
\bibliographystyle{IEEEbib}
\bibliography{refs}
\end{document}